\documentclass[12pt]{article}
\usepackage{graphicx,color}
\usepackage{hyperref} 

\usepackage{amssymb,amsfonts,amsmath,cancel,cite,multirow}
\usepackage[capitalise]{cleveref}

\newcommand{\ave}[1]{\left \langle #1 \right \rangle}

\newcommand{\pdiff}[3]{
\if 1#1   \frac{\partial #2 }{\partial #3 }
\else  \frac{\partial^#1#2 }{\partial #3^#1 } \fi}

\newcommand{\diff}[3]{
\if 1#1  \frac{{\rm d} #2 }{{\rm d} #3 }
\else  \frac{{\rm d}^{#1} #2 }{{\rm d}#3^{#1} } \fi
}         
                          
\newcommand{\Tr}[0]{{\rm Tr}}

\newcommand{\lam}{\lambda}

\newcommand{\ep}{\epsilon}
\newcommand{\del}{\delta}
\newcommand{\Del}{\Delta}
\newcommand{\ovl}{\overline}

\newcommand{\Si}{\Sigma}

\newcommand{\DRbar}[0]{$\ovl{\text{DR}}$ }

\newcommand{\eqs}[1]{\begin{equation}\begin{split} #1 \end{split}\end{equation}}

\begin{document}

\begin{titlepage}

\begin{flushright}
IPMU16-0026
\end{flushright}

\vskip 1.35cm
\begin{center}

{\large 
{\bf 
Threshold Corrections to Dimension-six Proton Decay Operators in Non-minimal SUSY $SU(5)$ GUTs
}}
\vskip 1.2cm

Borut Bajc$^{a}$,
Junji Hisano$^{b,c,d}$, 
Takumi Kuwahara$^c$
and
Yuji Omura$^b$\\

\vskip 0.4cm

{\it $^a$Jo\v{z}ef Stefan Institute, 1000 Ljubljana, Slovenia}\\
{\it $^b$Kobayashi-Maskawa Institute for the Origin of Particles and the Universe (KMI),
Nagoya University, Nagoya 464-8602, Japan}\\
{\it $^c$Department of Physics,
Nagoya University, Nagoya 464-8602, Japan}\\
{\it $^d$
Kavli IPMU (WPI), UTIAS, University of Tokyo, Kashiwa, Chiba
 277-8584, Japan}
\date{\today}

\vskip 1.5cm

\begin{abstract} 
We calculate the high and low scale threshold corrections to the $D=6$ proton decay mode in supersymmetric 
$SU(5)$ grand unified theories with higher-dimensional representation Higgs multiplets. In particular, we 
focus on a missing-partner model in which the grand unified group is spontaneously broken by the $\mathbf{ 75}$-dimensional 
Higgs multiplet and the doublet-triplet splitting problem is solved. We find that in the missing-partner model the $D=6$ proton decay rate gets suppressed 
by about $60\%$, mainly due to the threshold effect at the GUT scale, while the SUSY-scale threshold corrections are 
found to be less prominent when sfermions are heavy.

\end{abstract}

\end{center}
\end{titlepage}

\section{Introduction \label{sec:intro}}
Grand unified theories (GUTs) are attractive candidates for physics beyond the standard model (SM).
The unification of the SM gauge groups $SU(3)_C \times SU(2)_L \times U(1)_Y$ provides a unified 
description both of gauge interactions and of matter fields. Besides, supersymmetry (SUSY) indicates 
the precise unification of the SM gauge couplings at the energy scale $\sim 10^{16}~{\rm GeV}$ 
\cite{Dimopoulos:1981yj,Ibanez:1981yh,Einhorn:1981sx,Marciano:1981un} and provides a candidate 
for dark matter. The discovery of a scalar boson with the mass of $126~{\rm GeV}$ 
\cite{Aad:2012tfa,Chatrchyan:2012xdj,Aad:2015zhl} is consistent with the expectations for the SM Higgs boson. 
In spite of efforts to find deviations from the SM predictions and/or direct detection of the SUSY 
particles at the LHC run-2 experiments, no such signal has been found so far, see for instance ATLAS and 
CMS collaborations reports in Refs. \cite{Aad:2014kra,Aad:2014vma,Aad:2014wea,Chatrchyan:2013xna,Chatrchyan:2014lfa,CMS:2014dpa}.

Indirect measurements of rare decays or rare processes are useful to constrain new physics.
In particular, SUSY GUTs generically predict nucleon decays by exchanging an additional 
gauge boson called the $X$ boson ($D=6$ decay) or a color-triplet Higgs multiplet 
($D=5$ decay). In this paper we will 
assume that the $R$-parity violating ($D=4$ decay) mode is absent or at least negligible. 

Regarding the $D=5$ decay mode it is well-known that the minimal renormalizable SUSY $SU(5)$ GUT 
in the low-scale SUSY scenario (spartners typically around 1 TeV) has been excluded by predicting 
a too short lifetime \cite{Goto:1998qg,Murayama:2001ur}. Several ways to relax this severe 
constraint have been considered. For example, imposing the Peccei-Quinn (PQ) symmetry \cite{Peccei:1977ur}, one 
can suppress the baryon-number violating terms in the superpotential \cite{Hisano:1992ne,Hisano:1994fn}. 
Similarly, in the high-scale SUSY \cite{Hall:2009nd} or split SUSY scenario \cite{ArkaniHamed:2004fb,Giudice:2004tc,Ibe:2012hu}, large sfermion masses reduce the Wilson coefficients of the four-Fermi operators responsible for nucleon decay via the color-triplet Higgs multiplet \cite{Hisano:2013exa,Bajc:2015ita}. Another possible way to avoid a too fast proton decay rate is to use higher dimensional operators to increase the GUT and triplet scales 
\cite{Bachas:1995yt,Chkareuli:1998wi,Bajc:2002bv,Bajc:2002pg} and/or to suppress the color triplet Yukawa 
couplings without affecting the fermion masses \cite{Bajc:2002pg,EmmanuelCosta:2003pu}, or to assume some 
very specific flavor structure \cite{Bajc:2002bv}. Last but not least, 
models originating from higher spacetime dimensions can make use of continuum or discrete symmetries to completely or 
partially suppress the $D=5$ mode \cite{Hall:2001pg,Witten:2001bf,Klebanov:2003my,Friedmann:2002ty}. 
The bottom line is that the $D=5$ decay mode is potentially dangerous but very model dependent. 

An opposite situation is with the $D=6$ mode, which is on one side typically slower than the $D=5$ one, 
but on the other side more predictive, less model dependent. In this paper we will consider in detail this 
mode. The results will thus be particularly interesting for models in which for some reason the $D=5$ 
mode is negligible and the $D=6$ one dominates.

The main proton decay mode via the $D=6$ gauge interaction is into a neutral pion and a positron. For this 
decay mode various next-to-leading order corrections have been considered: the two-loop renormalization-group equations (RGEs) for the Wilson coefficients in MSSM \cite{Hisano:2013ege} and SM \cite{Daniel:1983ip}, 
below the electroweak (EW) scale \cite{Nihei:1994tx}, and the one-loop threshold corrections at the GUT scale 
in the minimal SUSY $SU(5)$ model \cite{Hisano:2015ala}. However, such corrections are not available for 
extended SUSY GUT models, which are motivated by the solution to the doublet-triplet splitting problem. 
These models typically employ large Higgs representations, so threshold corrections are expected 
to be particularly important. Since the anomalous dimensions include only gauge couplings at the next-to-leading 
order, we will focus in the following only on the threshold corrections by gauge interactions.

In this paper we will estimate the threshold effect in SUSY $SU(5)$ GUT models with negligible proton decay 
via the color-triplet Higgs exchange. As mentioned above, this can be easily obtained for example by imposing a global 
symmetry such as PQ symmetry. On top of that a light (below the GUT scale) color triplet is typically needed for exact 
unification \cite{Murayama:2001ur}. A complete model with both ingredients is the missing-partner model 
\cite{Grinstein:1982um,Masiero:1982fe}: it naturally solves the doublet-triplet splitting problem of the SUSY 
$SU(5)$ GUT since the $\mathbf{ 5} ~ (\ovl{\mathbf{5}})$ Higgs multiplet only couples to the $\ovl{\mathbf{50}} ~ (\mathbf{50})$ multiplet 
which however does not contain the doublet partner of $\mathbf{ 5} ~ (\ovl{\mathbf{ 5}})$. Since the adjoint $\mathbf{ 24}$ cannot 
couple $\mathbf{ 5}$ with $\ovl{\mathbf{50}}$ (it does couple $\mathbf{ 5}$ with $\ovl{\mathbf{45}}$ \cite{Zhang:2012rc} though), 
a higher dimensional adjoint $\mathbf{ 75}$ is needed. Then this Higgs multiplet spontaneously 
breaks the unified gauge symmetry into the SM gauge groups. 
As a result, there are a large number of fields contributing to the vacuum polarization of the $X$-boson supermultiplet 
and hence the threshold correction at the GUT scale may affect a precise predictions of proton lifetime. 
A similar analysis was carried out for the $D=5$ decay mode in the context of SUSY $SO(10)$ GUTs in
\cite{Aulakh:2013lxa}, where it was concluded that the huge number of fields involved sensibly affects the wave function renormalization factor.

The current status of the nucleon decay experiments is as follows: the partial lifetime limit on $p \to \pi^0 e^+$ is $\tau(p \to \pi^0 e^+) > 1.67 \times 10^{34}$ years, and the bound on the partial lifetime for $p \to K^+ \ovl\nu$ is $\tau(p \to K^+\ovl\nu) > 6.6 \times 10^{33}$ years \cite{Ikeda:2015pre,Takhistov:2016eqm}.
It is expected that a future experiment, the Hyper-Kamiokande, may achieve a sensitivity of 5-10 times the present bound.

This paper is organized as follows:
in \cref{sec:model}, we briefly introduce both the minimal and the missing-partner set-up of the renormalizable 
SUSY $SU(5)$ GUT. In \cref{sec:tc_gut}, we show the one-loop threshold corrections to the baryon-number 
violating dimension-six operators at the GUT scale. We estimate them at the GUT scale in the minimal SUSY 
$SU(5)$ and the missing-partner $SU(5)$ models, and then compare the numerical results for threshold effects 
in each models in \cref{sec:num_est}. Finally, we summarize this paper in \cref{sec:concl}. 
For completeness, in \cref{sec:tc_susy}, we derive threshold corrections at the SUSY scale and compute their 
numerical values. In \cref{sec:fielddcmp}, we list the threshold contributions from the irreducible representations 
with the Dynkin index below that of the $\mathbf{ 75}$-dimensional multiplet.
In \cref{sec:GUTvevint}, we give the interaction terms including the GUT-breaking vacuum expectation values (VEVs) in the cases of the adjoint Higgs multiplet and the $\mathbf{ 75}$-dimensional Higgs multiplet.

\section{Models \label{sec:model}}

To begin with, we briefly review the SUSY $SU(5)$ GUTs.
The matter supermultiplets are completely embedded in the $\ovl{\mathbf{5}} + \mathbf{ 10}$ representation for each generation:
\eqs{
\Phi_{iA}(\mathbf{\bar 5})=\left(
\begin{array}{c}
D^C_{i\alpha} \\
\ep_{rs} L^s_i
\end{array}\right), ~~~~~ 
\Psi^{[AB]}_i(\mathbf{10})= \frac{1}{\sqrt 2}\left(
\begin{array}{cc}
\ep^{\alpha\beta\gamma} U^C_{i\gamma} & Q^{r\alpha}_i \\
- Q^{s\beta}_i & \ep^{sr} E^C_i
\end{array}
\right),
}
where $A, B, \cdots = 1, 2, \cdots, 5$ are the $SU(5)$ indices, $\alpha, \beta, \cdots = 1, 2, 3$ and 
$r, s, \cdots = 1,2$ are the $SU(3)_C$ and $SU(2)_L$ indices, respectively. $ i = 1, 2, 3$ denotes 
the generation. The component superfields describe the chiral superfields of the minimal supersymmetric 
standard model (MSSM); $D^C, U^C$ and $E^C$ are the right-handed charge conjugated down-type, 
up-type, and charged lepton superfields, while $Q$ and $L$ denote the left-handed quark and lepton 
doublet superfields, respectively. A square bracket $[ \dots ]$ represents antisymmetric indices.

Differences among SUSY $SU(5)$ GUTs appear in the Higgs sector.
We describe the Higgs sector and the mass spectrum in the minimal and missing-partner models in the following 
two subsections.

\subsection{Higgs Sector in the Minimal $SU(5)$}

In the Higgs sector, two types of Higgs multiplets are required. One is the Higgs multiplet 
including the MSSM Higgs multiplets which are needed for the electroweak symmetry breaking.
The MSSM Higgs multiplets are embedded in the minimal $SU(5)$ into the $\mathbf{ 5} + \ovl{\mathbf{5}}$ 
(denoted by $H$ and $\ovl H$) and are so accompanied with the color-triplet Higgs multiplets. 
The other, additional, Higgs multiplet spontaneously breaks the GUT gauge group. 
In the minimal SUSY $SU(5)$ GUT, this role is played by the adjoint $\mathbf{ 24}$-dimensional Higgs 
 multiplet (denoted by $(\Si_{24})^A_B$), whose fields are either eaten by the heavy $X$ 
gauge boson (the would-be Nambu-Goldstone fields) or are typically heavy.

The superpotential in the Higgs sector depends on the particle content. As we said above, in the minimal 
setup of the SUSY $SU(5)$, the Higgs sector is only composed of the $\mathbf{ 5} + \ovl{\mathbf{5}}$ Higgs multiplets 
and the adjoint Higgs multiplet. The superpotential for the Higgs sector in the minimal renormalizable SUSY $SU(5)$ is
\eqs{
W_{\rm Minimal} = & ~ 
\frac{f}{3} \Tr (\Si_{24})^3 + \frac{m_{24}}{2} \Tr (\Si_{24})^2 + \lam \ovl H_A ((\Si_{24})^A_B + 3v_\mathbf{ 24} \delta^A_B) H^B.
}
Here, $v_\mathbf{ 24} = m_{24}/f$ denotes the VEV of the adjoint Higgs multiplet.
In the last term, we fine-tuned the parameters between $\ovl H \Si_{24} H$ and $\ovl H H$ so to get the MSSM Higgs 
doublets massless after symmetry breaking.

In the minimal SUSY $SU(5)$ the adjoint Higgs multiplet $(\Si_{24})^A_B$ ($A,B=1,\cdots,5$) obtains 
a GUT-breaking VEV. The K\"ahler potential for the adjoint Higgs multiplet is given by
\eqs{
\mathcal{K}_{\mathbf{24}} = (\Si_{24}^\dag)^A_B (e^{2g_5V})^B_C (e^{-2g_5V})^D_A (\Si_{24})^C_D,
}
where $g_5$ and $V$ denote the gauge coupling and the vector superfield in $SU(5)$.
Parametrizing the adjoint Higgs VEV as 
\eqs{
\ave{(\Si_{24})^r_s} = - 3 v_{\mathbf{24}} \del^r_s, ~~~~~
\ave{(\Si_{24})^\alpha_\beta} = 2 v_{\mathbf{24}} \del^\alpha_\beta, \label{eq:vev24}
}
the mass of the $X$ boson is given by $M_X = 5  g_5 v_{\mathbf{24}}$.

In this model, the adjoint Higgs multiplet is decomposed into the color-octet, the weak-triplet, the SM singlet, and the would-be 
Nambu-Goldstone multiplets. The octet and triplet have the same mass $M_{\Si_{24}} = 5 m_{24}/2$, while the SM singlet 
has the mass of $M_{\Si_{24}}/5$. The color triplets obtain the mass of $M_{H_C} = 5 \lam v_{\mathbf{24}}$ after symmetry 
breaking.

\subsection{Higgs Sector in the Missing-partner $SU(5)$}

In the missing-partner model, the MSSM Higgs doublets become massless without fine-tuning. 
In this model, the $\mathbf{ 50}+\ovl{\mathbf{50}}$-dimensional Higgs multiplets are introduced to give GUT-breaking 
mass to the color-triplet Higgs multiplets via couplings with $\mathbf{75}$-dimensional Higgs. This 
$\mathbf{ 75}$-dimensional Higgs multiplet includes the SM singlet \cite{Grinstein:1982um,Masiero:1982fe} 
and so can break $SU(5)$ to SM. In the following it is denoted by $ (\Si_{75})^{[AB]}_{[CD]}$. The dangerous 
$D=5$ proton decay operators are suppressed by imposing a PQ symmetry \cite{Hisano:1994fn}. 
We thus introduce in this model two $\mathbf{ 50}+\ovl{\mathbf{ 50}}$ pairs (denoted by 
$\Theta, \ovl \Theta, \Theta', \ovl \Theta'$) and an additional $\mathbf{ 5}+\ovl{\mathbf{5}}$ pair (denoted by $H', \ovl H'$).
The PQ charge assignment for them is given in \cref{tab:PQcharge}.
The invariant superpotential under the gauge and global symmetries is given as follows\footnote
{Similar to the original missing-partner model \cite{Grinstein:1982um,Masiero:1982fe}, the terms $H\ovl{H}'$ and 
$H'\ovl{H}$ are omitted by hand even if the PQ symmetry is imposed \cite{Hisano:1994fn}. In this paper we adopt this model as a prototype model with higher-dimensional representation Higgs multiplets.
}
:
\eqs{
W_{\rm MP} = & ~
g_H \ep_{ABCDE} H^A (\Si_{75})^{[BC]}_{[FG]} \Theta^{[DE][FG]}
+ g_{\ovl H} \ep^{ABCDE} \ovl H_A (\Si_{75})_{[BC]}^{[FG]} \ovl \Theta_{[DE][FG]} \\
& + g'_H \ep_{ABCDE} H'^A (\Si_{75})^{[BC]}_{[FG]} \Theta'^{[DE][FG]}
+ g'_{\ovl H} \ep^{ABCDE} \ovl H'_A (\Si_{75})_{[BC]}^{[FG]} \ovl \Theta'_{[DE][FG]} \\ 
& + m_{75} (\Si_{75})_{[AB]}^{[CD]} (\Si_{75})_{[CD]}^{[AB]}
- \frac13 \lam_{75} (\Si_{75})^{[AB]}_{[EF]} (\Si_{75})^{[CD]}_{[AB]} (\Si_{75})^{[EF]}_{[CD]} \\
& + M_1 \ovl \Theta_{[AB][CD]} \Theta'^{[AB][CD]} 
+ M_2 \ovl \Theta'_{[AB][CD]} \Theta^{[AB][CD]}. 
}
Already the large Dynkin index of $\mathbf{75}$ implies a asymptotically non-free theory above the GUT scale. 
Additional $\mathbf{ 50}+\ovl{\mathbf{50}}$ pairs would lead to the Landau pole between the GUT and Planck scales. 
By itself this may not be a problem, and could signal the presence of a non-perturbative UV fixed point 
\cite{Litim:2014uca,Litim:2015iea}, although in supersymmetry this may not be easy to obtain 
\cite{Intriligator:2015xxa}. To simplify the analysis we will assume in the following a perturbative GUT 
all the way to the Planck scale. This means that the masses of $\mathbf{ 50}+\ovl{\mathbf{50}}$ pairs must be at 
the gravitational scale ($M_1 = M_2 = M_{\rm Pl}$).

\begin{table}[t]
\begin{center}
\caption{Field contents in missing-partner model with Peccei-Quinn symmetry.}
\begin{tabular}{|l|c|c|c|c|c|c|c|c|c|c|c|c|c|}\hline
&$\Phi$&$\Psi$&$H$&$\ovl{H}$&
$H'$&$\ovl{H}'$&
$\Theta$&$\ovl{\Theta}$&
$\Theta'$&$\ovl{\Theta}'$&
$\Si_{75}$&$P$&$Q$\\
\hline\hline
dim. &$\mathbf{ \ovl{5}}$&$\mathbf{ 10}$&$\mathbf{ 5}$&$\mathbf{ \ovl{5}}$
&$\mathbf{ 5}$&$\mathbf{ \ovl{5}}$
&$\mathbf{ 50}$&$\mathbf{ \ovl{50}}$
&$\mathbf{ 50}$&$\mathbf{ \ovl{50}}$
&$\mathbf{ 75}$&$\mathbf{ 1}$&$\mathbf{ 1}$\\
\hline
$U(1)_{\rm PQ}$
&$\frac12$&$\frac32$&$-3$&$-2$&
$2$&$3$&
$3$&$2$&
$-2$&$-3$&$0$&$-5$&$15$\\
\hline
\end{tabular}
\label{tab:PQcharge}
\end{center}
\end{table}

In the missing-partner model with PQ symmetry it is the VEV of the $\mathbf{75}$ representation superfield 
which breaks the GUT symmetry. The K\"ahler potential for the $\mathbf{75}$ representation superfield is given by;
\eqs{
\mathcal{K}_{\mathbf{75}} = (\Si_{75}^\dag)^{[AB]}_{[CD]} (e^{2g_5 V})^C_E (e^{2g_5 V})^D_F (e^{-2g_5V})^G_A (e^{-2g_5V})^H_B  (\Si_{75})^{[EF]}_{[GH]}.
}
With the VEV of the $\mathbf{75}$ multiplet given by
\eqs{
\ave{(\Si_{75})^{[rs]}_{[tu]}} 
& = \frac32 v_{\mathbf{75}} (\del^r_t \del^s_u - \del^r_u \del^s_t ), ~~~~~
\ave{(\Si_{75})^{[\alpha \beta]}_{[\gamma \del]}} 
= \frac12 v_{\mathbf{75}} (\del^\alpha_\gamma \del^\beta_\delta - \del^\alpha_\delta \del^\beta_\gamma ), \\
\ave{(\Si_{75})^{[\alpha r]}_{[\beta s]}} 
& = - \frac12 v_{\mathbf{75}} \del^\alpha_\beta \del^r_s, \label{eq:vev75}
}
the mass of the $X$ boson equals $M_X = 2 \sqrt 6 g_5 v_{\mathbf{75}}$.
We also find easily $v_{\mathbf{75}} = 3m_{75}/2\lam_{75}$ by imposing $F$-term conditions.

\begin{table}[t]
\caption{Mass splitting in the $\mathbf{ 75}$-dimensional Higgs. $(r_C, r_W)_{Y}$ denotes the irreducible representation transforming as $SU(3)_C$ $r_C$-plet and $SU(2)_L$ $r_W$-plet with hypercharge $Y$ under the SM gauge groups.}
\begin{center}
\begingroup
\renewcommand{\arraystretch}{1.2}
\begin{tabular}{|c|c|}
irrep. & Mass \\ \hline
$(\mathbf{1},\mathbf{1})_{0} $ & $\frac25 M_{\Si_{75}}$ \\
$(\mathbf{3},\mathbf{1})_{-\frac53}, ~~
(\ovl{\mathbf{3}},\mathbf{1})_{\frac53}$ & $\frac45 M_{\Si_{75}} $\\
$(\mathbf{3},\mathbf{2})_{\frac56}, ~~
(\ovl{\mathbf{3}},\mathbf{2})_{-\frac56}$ & 0 ~~ (Nambu-Goldstone) \\
$ (\ovl{\mathbf{6}},\mathbf{2})_{\frac56}, ~~
(\mathbf{6},\mathbf{2})_{-\frac56} $ & $\frac25 M_{\Si_{75}}$ \\
$ (\mathbf{8},\mathbf{1})_{0} $ & $\frac15 M_{\Si_{75}}$ \\
$ (\mathbf{8},\mathbf{3})_{0} $ & $M_{\Si_{75}} \equiv 5 m_{75}$ \\
\end{tabular}
\endgroup
\end{center}
\label{tab:mass_75}
\end{table}%

The GUT-breaking VEV of the $\mathbf{ 75}$-dimensional Higgs multiplet also gives rise to the mass splitting among 
its components. The full spectrum is shown in \cref{tab:mass_75}. 
The color triplets $H_C, H_C', \ovl H_C$, and $\ovl H_C'$ obtain masses after integrating out two $\textbf{50} + \ovl{\mathbf{50}}$ pairs as follows,
\eqs{
W = M_{H_C} H_C \ovl H_C' + M_{H'_C} H_C' \ovl H_C, 
}
with masses defined as
\eqs{
M_{H_C} \equiv \frac{48 v_{\mathbf{75}}^2}{M_{\rm Pl}} g_H g_{\ovl H}', ~~~~ 
M_{\ovl H_C} \equiv \frac{48 v_{\mathbf{75}}^2}{M_{\rm Pl}} g_H' g_{\ovl H}.
}
Therefore, we have a relatively small mass  $M_{H_C}, M_{\ovl H_C} \sim 10^{15} ~ {\rm GeV}$ if we take the reduced Planck mass as $M_{\rm Pl} = 2.4 \times 10^{18}~{\rm GeV}$ and $g_H v_{\mathbf{75}}, g_H' v_{\mathbf{75}}, g_{\ovl H} v_{\mathbf{75}}, g_{\ovl H}' v_{\mathbf{75}} \sim 10^{16}~{\rm GeV}$.
On the other hand, there remain four massless $SU(2)_L$ doublets, $H_f, \ovl H_f, H_f'$, and $\ovl H_f'$ so far.

In order to break the PQ symmetry, we should introduce a pair of $SU(5)$ singlets $P$ and $Q$, which ensure that two Higgs doublets obtain the mass of the intermediate scale $\sim 10^{11}~{\rm GeV}$ \cite{Hisano:1994fn,Murayama:1992dj}.
In fact, in the original paper \cite{Hisano:1992ne} for the missing-partner model with PQ symmetry, a scalar potential for $P$ and $Q$ is induced by the following superpotential
\eqs{
W_{\rm PQ} = \frac{f_{\rm PQ}}{M_{\rm Pl} } P^3 Q + g_P \ovl H'_A H'^A P,
}
and the negative soft SUSY breaking mass $-m^2$ for $P$.
In this setup, VEVs for $P$ and $Q$ are given as 
\eqs{
\ave{P} \simeq \ave{Q} \simeq \sqrt{\frac{M_{\rm Pl} m}{f_{\rm PQ}}} \sim 10^{11}~ {\rm GeV},
}
where we assume $m \sim 1~{\rm TeV}$ and $f_{\rm PQ} \sim 1$.
The second term of $W_{\rm PQ}$ gives rise to the mass of a pair of two Higgs doublets ($M_{H_f'} = g_P \ave{P}$) and the dimension-five operator for proton decay.
However, the dimension-five operator is suppressed due to the effective color-triplet mass $M_{H_C}^{\rm eff} = M_{H_C} M_{\ovl H_C}/M_{H_f'}$.
\section{Threshold Correction at the GUT Scale \label{sec:tc_gut}}

In this section, we estimate the vacuum polarization function in the extended $SU(5)$ GUT models.
As mentioned in \cref{sec:intro}, we only focus on the gauge interaction since the calculation of only 
two-loop RGEs are carried out. The other threshold corrections, such as vertex and box corrections 
at the GUT scale, have been estimated in \cite{Hisano:2015ala}.
While vertex and box corrections are independent from the Higgs sector of the GUT scale, 
the vacuum polarization of $X$ boson is affected by them.
Before we focus on the vacuum polarization of the $X$-boson vector superfields, we summarize the threshold corrections at the GUT scale.

The effective K\"ahler potential for dimension-six operators is given by
\eqs{
{\cal L}_{\text{dim.6}} = & \int d^4 \theta \left( \sum_{i=1}^2 C^{(i)} {\cal O}^{(i)} + {\rm h.c.} \right)\label{eq:dim_6},
}
with operators ${\cal O}^{(i)}~(i=1,2)$ defined as
\eqs{
{\cal O}^{(1)} & = \ep_{\alpha\beta\gamma} \ep_{rs} U^{C \dag \alpha} D^{C\dag \beta}  Q^{r \, \gamma} L^s , ~~~~~
{\cal O}^{(2)} = \ep_{\alpha\beta\gamma} \ep_{rs} E^{C\dag} U^{C \, \dag \, \alpha} Q^{r\beta} Q^{s\gamma},
}
where we suppress the flavor indices. The Wilson coefficients $C_{\rm GUT}^{(i)}$ are defined as
\eqs{
C_{\rm GUT}^{(1)} & = C_{\rm GUT}^{(2)} = - \frac{g_5^2}{M_X^2}.
}

In matching the amplitudes of the low (MSSM) and high (GUT) energy effective theories, 
we include the threshold correction of the massive particles \cite{Hisano:2015ala}, 
\eqs{
C^{(i)} = (1-\lam_{\rm GUT}^{(i)}) C_{\rm GUT}^{(i)}
}
with
\eqs{
\lam^{(1)}_{\rm GUT} (\mu)
& = \frac{\Si_X(0)}{M_X^2+\Si_X(0)} + \frac{g_5^2}{16\pi^2} \frac{16}5 \left( 1 - \ln \frac{M_X^2}{\mu^2} \right), \\
\lam^{(2)}_{\rm GUT} (\mu)
& = \frac{\Si_X(0)}{M_X^2+\Si_X(0)} + \frac{g_5^2}{16\pi^2} \frac{18}5 \left( 1 - \ln \frac{M_X^2}{\mu^2} \right).
}
Here $M_X$ denotes the mass of the $X$ boson and $\Si_X(0)$ is the correction to the $X$-boson mass. 
The first term for each $\lam_{\rm GUT}^{(i)}$ arises from the vacuum polarization of the $X$ boson, while the second 
term describes the correction from box and vertex diagrams.

The renormalized two-point function for the $X$ boson is given by 
\eqs{
\Gamma^{(2)}_X(p^2) = p^2 - M_X^2 - \Si_X(p^2),
}
where $p^2$ is the $X$-boson momentum square. Here we adopt the on-shell scheme for the $X$-boson mass, 
\eqs{
\Si_X(p^2) = \ovl\Si_X(p^2) - \ovl\Si_X(M_X^2),
} 
while we use the \DRbar scheme otherwise. Particles with the $SU(5)$ invariant mass much heavier 
than the $X$-boson mass get decoupled from $\Si_X(0)$ under the on-shell scheme. 

\begin{figure}[t]
\begin{center}
\includegraphics[width=7cm,clip]{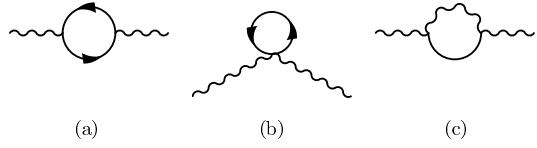}
\caption{
Radiative corrections due to chiral multiplets to two-point function of the $X$ superfield. 
Solid and wavy lines correspond to chiral superfields and vector superfields, respectively.
}
\label{fig:quad}
\end{center}
\end{figure}

The three diagrams in \cref{fig:quad} contribute to the radiative corrections to the $X$-vector multiplet 
two-point function from the (massive) chiral superfields. After picking the transverse mode and 
regularizing the UV divergence, we obtain the finite correction to the two-point function from the 
diagram (a) in \cref{fig:quad} as follows:
\eqs{
\Gamma_{XX}^{\text{(a)}} 
= \frac{g_5^2 b_{ij}}{16\pi^2} B(p^2,M_i^2,M_j^2) \int d^4 \theta X^{\dag \alpha}_r(-p,\theta) P_T X_\alpha^r(p,\theta)+\text{(longitudinal mode)}.
}
Here, $X(p,\theta)$ is a vector superfield including the $X$ boson and $\theta$ corresponds to the superspace 
Grassmann variable, while $P_T$ denotes the projection operator on the transverse mode in the 
superspace formulation. $b_{ij}$ denotes the group-theoretical factor, which depends on the representation of internal 
chiral superfields. They are tabulated in \cref{sec:fielddcmp} for several lower dimensional representations. 
Finally
\eqs{
B(p^2,M_i^2,M_j^2) & \equiv \int^1_0 dx \left[ \Delta- (2\Delta + x(x-1)p^2 ) \ln \frac{\Delta}{\mu^2} \right], 
\label{eq:loop_func1}
}
where 
\eqs{
\Delta = x(x-1) p^2 + x M_j^2 + (1-x) M_i^2
\label{eq:Delta}
}
and $M_{i,j}$ are the masses of the chiral superfields in the 
loop diagram, while $\mu$ is the renormalization scale.

The contribution of the diagram (b) in \cref{fig:quad}, $\Gamma^{\text{(b)}}_{XX}$, is not vanishing.
However, since $\Gamma^{\text{(b)}}_{XX}$ has no $p^2$-dependence and we take the on-mass shell scheme 
for the $X$-boson mass, this does not contribute to $\Si_X(p^2)$.

The third contribution (the diagram (c) in \cref{fig:quad}) comes from the interactions which include the VEV of the 
GUT-breaking Higgs superfield:
\eqs{
\Gamma_{XX}^{\text{(c)}} 
= \frac{g_5^2 M_X^2 a_{ij} }{16\pi^2} A(p^2,M_i^2,M_j^2) \int d^4 \theta X^{\dag \alpha}_r(-p,\theta) P_T X_\alpha^r(p,\theta).
}
Here, $a_{ij}$ is also a group-theoretical factor, similar as $b_{ij}$, and 
\eqs{
A(p^2,M_i^2,M_j^2) & \equiv \int^1_0 dx  \ln \frac{\Delta}{\mu^2}, \\
}
with $\Delta$ given in \cref{eq:Delta}.

Let us now discuss the contributions to the vacuum polarization of the $X$-boson vector superfield coming 
from different representations. First, the fields in the irreducible representation of $SU(5)$ are decomposed 
into irreducible representations of the SM gauge groups. We give the SM decomposition of some GUT multiplets 
in \cref{sec:fielddcmp}. The vacuum polarization of the $X$ boson is given by 
\eqs{
\ovl \Si^{\mathrm{Rep.}}_X (p^2) 
= \frac{g_5^2}{16\pi^2} \sum_{i,j} b_{ij} B(p^2,M_i^2,M_j^2)
+ \frac{g_5^2 M_X^2 }{16\pi^2} \sum_{i} a_i A(p^2,M_X^2,M_i^2), \label{eq:vp_X}
}
where the superscript ``Rep.'' indicates the $SU(5)$ representation, such as $\mathbf{ 5}+\ovl{\mathbf{5}}$, $\mathbf{10}+\ovl{\mathbf{10}}$, 
$\mathbf{24}$, and so on. $i,j = 1, \cdots, N$ ($i,j=\ovl 1, \cdots, \ovl N$) denote the labels of irreducible 
representations of the SM gauge groups (and its complex conjugated representation). We neglect the $p^2$-independent 
terms since we take the on-mass shell condition for the vacuum polarization of the $X$-boson superfield. 
The vacuum polarization coefficients $a_i$ and $b_{ij}$ are determined by the interactions between the 
$X$-boson superfield and the corresponding chiral superfields.

The first term in \cref{eq:vp_X} comes from the gauge interaction between the $X$-boson and (anti-)chiral superfields.
The mass eigenvalues of the chiral superfields in the loop are denoted by $M_i$. We show the vacuum polarization 
coefficients $b_{ij}$ in \cref{tab:vp_X1} for some $SU(5)$ representations. The $b_{ij}$ not listed in \cref{tab:vp_X1} 
are zero. Note that the missing-partner model \cite{Hisano:1994fn} includes $\mathbf{50} + \ovl{\mathbf{50}}$ pairs to induce 
the masses of color-triplet Higgs multiplets. However, these multiplets do not contribute to the vacuum polarization 
of the $X$-boson since we assume their vector-like mass to be at the Planck scale in order to keep the theory in the 
perturbative regime. Here, we list $b_{ij}$ from $\mathbf{5}+ \ovl{\mathbf{5}}$, $\mathbf{ 24}$, and $\mathbf{75}$ representations.
In \cref{sec:fielddcmp}, we display $b_{ij}$ only from $SU(5)$ representations, 
whose Dynkin indices are smaller than the Dynkin index of the $\mathbf{75}$ representation.


\begin{table}[t]
\caption{Vacuum polarization coefficients $b_{ij}$ and $a_i$.
The coefficients which are not listed here are zero.
$a_0$ is the coefficient from the NG and MSSM vector supermultiplets loops,
while $a_i ~ (i=1,2,\cdots)$ is the coefficient from the massive vector and chiral supermultiplets loops.
 $b_{ij}$ and $a_{i}$ which are not listed here are zero.}
\begin{center}
\begingroup
\renewcommand{\arraystretch}{1.2}
\begin{tabular}[b]{|c|c|c|}
reps. & $b_{ij}$ & $a_i$ \\ \hline
$\mathbf{5}+\ovl{\mathbf{5}}$ 
& $b_{12} = b_{\ovl1 \ovl2} = 1$ & \\ \hline
\multirow{4}{*}{$\mathbf{24}$} 
&&$a_0 = 5/2$ \\
& $b_{14} = b_{15} = 3/2$ & $a_1 = 3$ \\ 
& $b_{24} = b_{25} = 8/3$ & $a_2 = 16/3$\\
& $b_{34} = b_{35} = 5/6$ & $a_3 = 5/3$ \\ \hline
\multirow{7}{*}{$\mathbf{75}$} 
& $b_{14} = b_{23} = 1/3$ & $a_0 = 5/2$ \\ 
& $b_{16} = b_{25} = 2$ & $a_1 = a_2 = 6$ \\ 
& $b_{37} = b_{47} = 4/3$ & $a_7 = 8/3$ \\
& $b_{38} = b_{48} = 2/3$ & $a_8 = 4/3$\\
& $b_{39} = b_{49} = 26/3$ & $a_9 = 12$ \\
& $b_{58} = b_{68} = b_{59} = b_{69} = 6$ &\\ 
\end{tabular}
\endgroup
\end{center}

\label{tab:vp_X1}
\end{table}%

The SM decomposition of $\mathbf{5}, \mathbf{ 24} $, and $\mathbf{ 75}$ representations is given by
\eqs{
\mathbf{5} 
& = \phi_1 (\mathbf{1},\mathbf{2})_{\frac12} \oplus \phi_2 (\mathbf{3},\mathbf{1})_{-\frac13}, \\
\mathbf{24} 
& = \phi_1(\mathbf{1},\mathbf{3})_0 \oplus \phi_2 (\mathbf{8},\mathbf{1})_0 
\oplus \phi_3 (\mathbf{1},\mathbf{1})_ 0 \oplus \phi_4 (\mathbf{3},\mathbf{2})_{-\frac56} 
\oplus \phi_5 (\ovl{\mathbf{3}},\mathbf{2})_{\frac56}, \\
\mathbf{75} 
& = \phi_1 (\ovl{\mathbf{3}},\mathbf{1})_{-\frac53} \oplus \phi_2 (\mathbf{3},\mathbf{1})_{\frac53} 
\oplus \phi_3 (\mathbf{3},\mathbf{2})_{-\frac56} \oplus \phi_4 (\ovl{\mathbf{3}},\mathbf{2})_{\frac56} 
\oplus \phi_5 (\mathbf{6},\mathbf{2})_{\frac56} \\
& ~~~ \oplus \phi_6 (\ovl{\mathbf{6}},\mathbf{2})_{-\frac56} \oplus \phi_7 (\mathbf{1},\mathbf{1})_0 
\oplus \phi_8 (\mathbf{8},\mathbf{1}) _0\oplus \phi_9 (\mathbf{8},\mathbf{3})_0.
}
Here, the subscripts of $\phi_i$ correspond to the labels in \cref{eq:vp_X}.

The second term in \cref{eq:vp_X} arises from the interactions involving the GUT-breaking VEV. 
Thus, $a_i = 0$ is satisfied if there is no SM singlet component in the irreducible representation of $SU(5)$. 
We choose the Feynman-t' Hooft gauge throughout this paper. Then the Nambu-Goldstone superfields 
have the same mass as the $X$ boson. Since the internal lines in the diagram (c) of \cref{fig:quad} include 
either $X$-boson vector superfields or the Nambu-Goldstone chiral superfields, the loop function $A$ 
always has $M_X$-dependence.

The interaction terms with the GUT-breaking VEV $v_{\mathbf{24}}$ or $v_{\mathbf{75}}$ are given in 
\cref{sec:GUTvevint}. In \cref{tab:vp_X1}, the coefficients $a_i$ from the $\mathbf{24}$ and $\mathbf{75}$ 
chiral superfields are listed. The $a_i$ not listed in \cref{tab:vp_X1} are zero. $a_0$ corresponds to the 
contribution from the Nambu-Goldstone (NG) supermultiplets and MSSM vector supermultiplets, 
while $a_i ~ (i=1,2,\cdots)$ arise from the massive vector supermultiplet and chiral supermultiplet loop diagrams.
Since the MSSM vector superfields are massless, $M_0 = 0$ is satisfied.

\section{Numerical Results \label{sec:num_est}}

Let us now estimate the threshold effects on the proton lifetime. We define the ratio of the proton decay rates as
\eqs{
R \equiv \frac{\left. A_S^{(1)2}+(1+|V_{ud}|^2)^2 A_S^{(2)2}\right|_{\rm w}}
{\left. A_S^{(1)2}+(1+|V_{ud}|^2)^2 A_S^{(2)2}\right|_{\rm w/o}},
\label{eq:ratio}
}
where subscripts w and w/o denote the decay rates with and without the threshold correction, respectively.
$A_S^{(i)}~(i=1,2)$ indicate the quantum corrections to the Wilson coefficients from RGEs between the 
GUT and EW scales as well as finite corrections:
\eqs{
A_S^{(i)} = (1-\lam_C^{(i)}) \frac{C^{(i)}(m_Z)}{C^{(i)}(M_{\rm GUT})}. \label{eq:as}
}
Here $\lam_C^{(i)}$ denotes the sum of threshold corrections to the Wilson coefficients $C^{(i)}$ at the scales where the heavy particles are integrated out.

For completeness we need to include the threshold correction at the SUSY scale. The explicit formulae for the 
SUSY threshold correction are given in \cref{sec:tc_susy}. We find that their effect is 
vanishing if sparticles are degenerate in mass, but even with a large mass hierarchy between 
sfermions and gauginos it is only about a few percent.

Let us consider now the effect of the additional fields at the GUT scale. The precise prediction for proton lifetime 
depends on the unified coupling and the mass spectrum at the GUT scale. In particular, the GUT mass spectrum 
and the value of the unified gauge coupling are connected through the threshold corrections to the gauge couplings 
at the GUT scale \cite{Hisano:1992mh,Hisano:1992jj}. Thus by requiring gauge coupling unification we get a constraint on the mass 
spectrum of the GUT particles through threshold corrections. At the GUT scale, the one-loop matching conditions 
for gauge couplings are given as
\eqs{
\frac{1}{g_i^2{(\mu)}} = \frac{1}{g_5^2{(\mu)}} - \lam_i (\mu), \label{eq:gauge_threshold}
}
where $g_i ~ (i = 1,2,3)$ are the SM gauge couplings and $\lam_i ~ (i=1,2,3)$ denote threshold corrections 
for gauge couplings. The threshold corrections $\lam_i$ depend on the details of the GUT particles mass 
spectrum. For the minimal renormalizable SUSY $SU(5)$ GUT, the mass of color-triplet Higgs multiplets 
$M_{H_C}$ and the combination of the mass parameters $M_X^2 M_{\Si_{24}}$ are determined by \cite{Hisano:1992mh,Hisano:1992jj}
\begin{eqnarray}
\label{rge1}
\frac{3}{g_2^2(\mu)} - \frac{2}{g_3^2(\mu)} - \frac{1}{g_1^2(\mu)}
& = & \frac{1}{8\pi^2} \frac{12}{5} \ln \frac{M_{H_C}}\mu,  \\
\label{rge2}
\frac{5}{g_1^2(\mu)} - \frac{3}{g_2^2(\mu)} - \frac{2}{g_3^2(\mu)}
& = & \frac{1}{8\pi^2} 12 \ln \frac{M_X^2 M_{\Si_{24}}}{\mu^3}. 
\end{eqnarray}
The right-hand sides of Eqs.~(\ref{rge1}) and (\ref{rge2}) are thus determined by the low-energy gauge 
couplings and the SUSY mass spectrum. Notice that $M_X$, which enters in the expression for the $D=6$ 
proton lifetime, cannot be determined; Eq. (\ref{rge2}) constrains only the combination $M_X^2M_{\Si_{24}}$. 
We treat the mass of the $X$ boson as a free parameter in the following numerical calculation, 
and we estimate the gauge couplings at the matching scale (i.e. GUT scale), which we take 
$\mu=2\times10^{16}$ GeV, by using the two-loop RGEs for gauge couplings. 
After we fix $M_X$, we determine the mass of the color triplet $M_{H_C}$ from Eq.~(\ref{rge1}) 
and that of the adjoint Higgs $M_{\Si_{24}}$ from Eq.~(\ref{rge2}).
One of Eqs.~(\ref{eq:gauge_threshold}) (for example $g_1$) is then used to get 
$g_5$ at the GUT scale.

In the missing-partner $SU(5)$ model, the combinations of the GUT masses are constrained as \cite{Hisano:1994fn};
\begin{eqnarray}
\label{rgemp1}
\frac{3}{g_2^2(\mu)} - \frac{2}{g_3^2(\mu)} - \frac{1}{g_1^2(\mu)}
& = &\frac{1}{8\pi^2} \left(\frac{12}{5} \ln \frac{M_{H_C}M_{\ovl H_C}}{M_{H_f'}\mu}
+ 6 \ln \frac{2^6}{5^5} \right),  \\
\label{rgemp2}
\frac{5}{g_1^2(\mu)} - \frac{3}{g_2^2(\mu)} - \frac{2}{g_3^2(\mu)}
& = &\frac{1}{8\pi^2} \left(12 \ln \frac{M_X^2 M_{\Si_{75}}}{\mu^3} + 54 \ln \frac54 \right). 
\end{eqnarray}
Here, the parameters $M_{H_C}$ and $M_{\ovl H_C}$ correspond to the mass of the color-triplet Higgs multiplets, 
while $M_{H_f'}$ denotes the mass of the extra Higgs doublet induced by the breaking of the PQ symmetry. 
$M_{\Si_{75}}$ is defined as the mass of the component fields $(\mathbf{8},\mathbf{3})_{0} $ in \cref{tab:mass_75}.
The constant terms arise from the mass splitting of the component fields of the $\mathbf{ 75}$-dimensional 
Higgs multiplet as we have shown in \cref{tab:mass_75}.

In the following analysis for the missing-partner model, we determine the combinations 
$M_{H_C} M_{\ovl H_C}/M_{H_f'}$ and $M_X^2 M_{\Si_{75}}$ from Eqs.~(\ref{rgemp1})-(\ref{rgemp2}). 
As in the case of the minimal $SU(5)$, for a given sparticle mass spectrum, $M_X$ cannot 
be determined: Eq.~(\ref{rgemp2}) gives only a relation between $M_X$ and $M_{\Si_{75}}$. 
We also define the unified coupling $g_5$ as in the minimal $SU(5)$. 
For simplicity, we take the typical mass scale for the color-triplets, $M_{H_C} = M_{\ovl H_C} = 10^{15}~{\rm GeV}$, 
as shown in \cref{sec:model}, so that $M_{H_f'}$ is given 
by Eq.~(\ref{rgemp1}) at the matching scale $\mu=M_{\rm GUT}=2\times10^{16}$ GeV. 

\begin{table}[t]
\caption{Threshold effects on the partial proton decay rate.
For simplicity, we assume that all superparticles are degenerate in mass $M_S = 1~{\rm TeV}$.}
\begin{center}
\begin{tabular}{|c|c|c|c|c|}
& \multicolumn{2}{|c|}{Minimal $SU(5)$}
& \multicolumn{2}{|c|}{Missing-Partner} \\ \hline
$M_X$ & $1.0\times 10^{16}~{\rm GeV}$ & $2.0\times 10^{16}~{\rm GeV}$
& $1.0\times 10^{16}~{\rm GeV}$ & $2.0\times 10^{16}~{\rm GeV}$ \\ \hline 
$A_S^{(1)}$ & 2.070 & 1.968 & 1.301 & 1.269\\
$A_S^{(2)}$ & 2.162 & 2.059 & 1.352 & 1.295\\
$R$ &1.10& 0.994
&0.429& 0.394 \\
$g_5$ &0.697&0.713&0.938&1.198 \\
$\tau(p \to e^+\pi^0)$ [years] &$1.38\times10^{35}$&$2.23\times10^{36}$
&$1.08\times10^{35}$&$7.09\times10^{35}$
\end{tabular}
\end{center}
\label{tab:list_threshold}
\end{table}%

We list the numerical results in the minimal SUSY $SU(5)$ and the missing-partner model 
in \cref{tab:list_threshold}. Here, we take the mass of all SUSY particles to be $1~{\rm TeV}$, 
for simplicity. Since proton lifetime strongly depends on the mass of the $X$ boson, 
we display the results for two choices of $M_X=1.0 \times 10^{16}~{\rm GeV}$ and 
$M_X=2.0 \times 10^{16}~{\rm GeV}$. 
By using the central values for gauge couplings at the EW scale,
the mass parameters determined by Eqs.~(\ref{rge1})-(\ref{rgemp2}) are obtained as follows:
\eqs{
M_{H_C} = 6.35 \times 10^{15}~\text{GeV}, ~~~~~~
(M_X^2 M_{\Si_{24}})^{1/3} = 1.48 \times 10^{16} ~ \text{GeV},
}
for the minimal SUSY $SU(5)$, 
\eqs{
\frac{M_{H_C}M_{\ovl H_C}}{M_{H_f'}} = 1.06\times10^{20} ~ \text{GeV}, ~~~~~~
(M_X^2 M_{\Si_{75}})^{1/3} = 5.43 \times10^{15} ~\text{GeV}
}
for the missing-partner model.
The quantities $A_S^{(i)} ~ (i = 1,2) $ in \cref{tab:list_threshold} 
show the short-range renormalization factors with the threshold corrections defined by \cref{eq:as}.
For each $M_X$, we get the threshold corrections in the minimal $SU(5)$ model as
\eqs{
\lam^{(1)} & = -4.94 \times10^{-2}, ~~~ 
\lam^{(2)} = -4.65 \times10^{-2}, ~~~~~~~
(\text{for } ~ M_X = 1.0 \times 10^{16}~{\rm GeV}), \\
\lam^{(1)} & = 1.98 \times10^{-3}, ~~~ 
\lam^{(2)} = 3.26 \times10^{-3}, ~~~~~~~
(\text{for } ~ M_X = 2.0 \times 10^{16}~{\rm GeV}), \\
}
and in the missing-partner model as
\eqs{
\lam^{(1)} & = 0.340, ~~~ 
\lam^{(2)} = 0.346, ~~~~~~~
(\text{for } ~ M_X = 1.0 \times 10^{16}~{\rm GeV}), \\
\lam^{(1)} & = 0.369, ~~~ 
\lam^{(2)} = 0.373, ~~~~~~~
(\text{for } ~ M_X = 2.0 \times 10^{16}~{\rm GeV}). \\
}

When we estimate the partial proton lifetime in each model, we use the proton decay matrix elements 
calculated by the lattice simulation at 2 GeV \cite{Aoki:2013yxa}. Note that the unified coupling $g_5$ 
in the missing-partner model is larger than the one in the minimal $SU(5)$. This is due to the following two reasons: 
1) the combination $M_X^2 M_{\Si_{75}}$ in the missing-partner model is slightly smaller than in the minimal 
$SU(5)$ due to the constant term present in the right-hand side of Eq. (\ref{rgemp2}) but not of Eq. (\ref{rgemp1}), 
and, 2) there are many components of $\mathbf{75}$ contributing 
to the threshold correction for the gauge couplings.

In Table \ref{tab:list_threshold} we see that threshold effects are negligible in the minimal SU(5) model 
($R\sim1$), but suppress the proton decay rate by an approximate factor $0.4$ in the missing-partner 
model. A much bigger effect comes from the variation of the $X$-boson mass $M_X$, and is unfortunately 
not under control. This change of the lifetime $\tau$ with $M_X$ in \cref{tab:list_threshold} 
can be understood by the approximate tree level relation $\tau\propto(M_X/g_5)^4$.

The threshold corrections from vertex and box contributions depend only on $M_X$ and $g_5$ in the context 
of the SUSY $SU(5)$ GUTs, and these values are of order $10^{-2}$ \cite{Hisano:2015ala}. Let us consider 
the model dependence of the threshold effect, which appears in the vacuum polarization of the $X$ boson. When
the heavy spectrum is degenerate, the dominant (first term) contribution to the vacuum polarization 
in Eq. (\ref{eq:vp_X}) is proportional to the one-loop beta function for the unified gauge coupling. 
In the minimal SUSY $SU(5)$, the contribution from the gauge supermultiplet dominates the 
vacuum polarization. On the other hand, since there exist many massive fields 
in the missing-partner model, the contribution from the chiral supermultiplets is enhanced, and the vacuum 
polarization is dominated by the chiral supermultiplets in the model. Therefore, the resulting threshold effects vary 
among GUT models due to the relative sign between the contributions from the gauge and the chiral 
supermultiplets. 
Notice that we include other contributions, such as vertex contributions, box contributions, 
and vacuum polarizations which arise from interactions including GUT breaking VEV, 
though they are subdominant in the missing partner model.

If we estimate the masses of the Higgs multiplets by using Eqs.~(\ref{rgemp1})-(\ref{rgemp2}) with $M_X$ lying around 
$10^{16}~{\rm GeV}$, all components can be lighter than the $X$ boson in the missing-partner model. The 
$\mathbf{75}$-dimensional Higgs multiplet has a number of component fields with mass different from the 
matching (GUT) scale, and thus there can be a large contribution of the threshold correction to the proton lifetime. 
As a result, the proton decay rate in the missing-partner 
model is suppressed about $60 \%$ with $M_X = 2.0 \times 10^{16}~{\rm GeV}$ as we show in 
\cref{tab:list_threshold}.

\section{Conclusion and Discussion\label{sec:concl}}

In this study, we have evaluated the threshold corrections to the proton decay operators giving rise to 
$p \to \pi^0 e^+$, especially those induced by the higher-dimensional representation Higgs multiplets. 
The structure of the Higgs sector in the extended GUT models appears in the vacuum polarization of 
the $X$ boson. As the threshold corrections depend on the choice of the high-energy model, we have 
focused on the missing-partner model which solves the doublet-triplet problem in $SU(5)$. In this 
model, the Higgs sector effectively contains the $\mathbf{75}$-dimensional Higgs multiplet and two 
$\mathbf{5}+\ovl{\mathbf{5}}$ pairs below the Planck scale. We have determined the mass spectrum at the GUT scale 
according to the low energy gauge coupling constants, fixing some mass parameters such as the $X$-boson mass. 
In such a case, many multiplets acquire masses around $10^{15}$ GeV through the GUT symmetry breaking.
Since the size of the unified gauge coupling is determined including threshold corrections, these mass differences 
make the coupling large. As a result,  due to the small masses and the large coupling, the contributions 
of all these multiplets sum up to a correction of about 60 \% to the proton lifetime.

We have not considered the threshold corrections from Yukawa couplings in this study. This is because 
no two-loop RGE analysis including Yukawa couplings for the baryon-number violating operators is 
available. Although this effect is not expected to be sizable due to the smallness of the Yukawa 
(except the top Yukawa), it should be included sooner or later to complete the two-loop level analysis.

In our analysis we also ignore the Planck-suppressed operators.
We assume $\mathbf{50}+\ovl{\mathbf{50}}$ pairs have the mass around the Planck-scale.
There might be higher-dimensional operators with Planck scale suppression in general.
The presence of such operators induces, via an operator 
$\Si_{75} \mathcal{W}^\alpha \mathcal{W}_\alpha/M_{\mathrm{Pl}}$, 
the correction to the gauge coupling $g_5^2$ of about a few percent after symmetry breaking. 
This correction affects also the determination of the GUT mass spectrum via gauge coupling unification constraints, 
which strongly depends on the coupling of the higher-dimensional operator.
Such effects of the Planck scale physics should be taken into account 
once the full UV description of the missing-partner model is available.

Finally, we notice remaining uncertainties in the precise determination of the proton lifetime.
The matrix elements of $p \to \pi^0 e^+$ is evaluated with lattice QCD, and they have at present around 30\% uncertainty \cite{Aoki:2013yxa}. We hope that this uncertainty will be reduced in future studies.

\section*{Acknowledgments}
This work is supported by Grant-in-Aid for Scientific research
from the Ministry of Education, Science, Sports, and Culture (MEXT),
Japan, No. 23104011 (for J.H. and Y.O.). 
The work of J.H. is also supported by World Premier
International Research Center Initiative (WPI Initiative), MEXT,
Japan. The work of B.B. is partially supported by the Slovenian Research Agency. 
B.B. thanks KMI and the Department of Physics of Nagoya University for hospitality.

\newpage
\appendix
\section*{Appendix}
\appendix
\section{Threshold Correction at the SUSY Scale\label{sec:tc_susy}}

In this appendix, we derive the threshold corrections of the dimension-six operators at the SUSY scale. 
We focus only on gauge interactions.

\subsection*{Two-Point Function}
Now, we estimate the threshold corrections to the two-point function of fermions. After picking the UV 
divergence, we obtain the two-point function of SM fermions $\psi$ including the one-loop finite 
corrections as follows:
\eqs{
i\Gamma^\psi_{\rm full} = \left[ 1 - \frac{1}{16\pi^2} \sum_a g_a^2 C_a(N) f(M_a^2, m_\phi^2) 
+ \text{(SM contributions)}\right] i \Gamma_0^\psi.
}
$\Gamma_0^\psi\equiv\slash{\hskip -2mm p}-m_\psi$ denotes the tree-level two-point function of $\psi$ with 
$p^\mu$ the four-momentum and $m_\psi$ the tree level mass. $M_a$ and $m_\phi$ are the 
masses of the gaugino and the superpartners of $\psi$, respectively. The second term arises from the 
gaugino-sfermion loop and the third term describes the contribution from the SM loops. $g_a$ and 
$C_a(N)$ are the SM gauge coupling and the Casimir invariant, respectively. For the $SU(N)$ gauge 
theory, $C_a(N) = (N^2-1)/2N$, and for the $U(1)_Y$ gauge theory, $C_a(N) = Y^2$. The loop function 
$f$ is defined as
\eqs{
f(x,y) \equiv 2 \int^1_0 ds (1-s) \ln \left[ s x+ (1-s) y \right]. \label{eq:loopf_susy}
}
Here, all mass parameters are normalized by $\mu$ which denotes the renormalization scale.
It is easily found that $f(x,x)=\frac12 \ln x$, which corresponds to the case of degenerate masses.

In the SM, the one-loop corrected two-point function has the form 
\eqs{
i\Gamma^\psi_{\rm SM} = \left[ 1 - \lam_\psi + \text{(SM contributions)}\right] i \Gamma_0^\psi,
}
where $\lam_\psi$ denotes the one-loop threshold correction to the two-point function of $\psi$.
Then, we get $\lam_\psi$ after matching the two-point functions in the two theories:
\eqs{
\lam_\psi = \frac{1}{16\pi^2} \sum_a g_a^2 C_a(N) f(M_a^2, m_\phi^2). \label{eq:lam_psi}
}

\subsection*{Four-Fermi Vertices}

The baryon-number violating dimension-six operators are induced by the non-renormalizable K\"ahler potential.
The supersymmetric interaction term including four-Fermi operators is given by;
\eqs{
C^{ijkl} \int d^4 \theta \Phi_i^\dag \Phi_j^\dag\Phi_k\Phi_l
& = - \frac{C^{ijkl} }{4} \left[ \square(\phi_i^\ast \phi_j^\ast) \phi_k \phi_l - 2 \partial_\mu (\phi_i^\ast \phi_j^\ast) \partial^\mu(\phi_k \phi_l)  + \phi_i^\ast \phi_j^\ast \square(\phi_k \phi_l) \right] \\
& - \frac{iC^{ijkl} }{2} (\phi_i^\ast \ovl\Psi_j + \ovl \Psi_i \phi_j^\ast) \gamma^\mu \overleftrightarrow{\partial_\mu} (\phi_k \Psi_l + \Psi_k \phi_l) \\
& - \frac{C^{ijkl} }{2} \ovl \Psi_i \gamma^\mu P_L \Psi_l \ovl \Psi_j \gamma_\mu P_L \Psi_k. \label{eq:d6op}
}
Here, $\Phi_i$ ($\Phi_i^\dag$) is a chiral (anti-chiral) superfield.
Scalar and four-component spinor components of $\Phi_i$ are denoted as $\phi_i$ and $\Psi_i$, respectively.
The symbols $\square$ and $\overleftrightarrow{\partial_\mu}$ are respectively defined as $\square = \partial^\mu \partial_\mu$ and $A\overleftrightarrow{\partial_\mu}B = A\partial_\mu B - \partial_\mu A B$.
The roman indices $i,j,k,l$ describe the gauge and flavor indices.
$C^{ijkl}$ denotes the Wilson coefficient of the operators.

\begin{figure}[t]
\begin{center}
\includegraphics[width=6cm,clip]{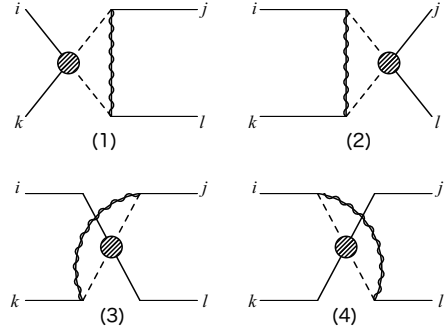}
\caption{Four-Fermi interactions induced by one-loop diagrams.
Blobs denote effective vertices including two fermions and two scalars, which are induced by the same K\"ahler potential.
Solid and dashed lines describe fermion and scalar lines, respectively, while wavy-solid lines indicate gauginos.
}
\label{fig:SUSYthreshold}
\end{center}
\end{figure}

\cref{fig:SUSYthreshold} shows the one-loop diagrams giving the four-Fermi interactions after integrating out superpartners.
In these figures, the blobs denote the interaction given by the second line of \cref{eq:d6op}.
Only four figures in \cref{fig:SUSYthreshold} contribute as finite corrections to nucleon decay matrix elements at the SUSY scale since one of two scalar fields must originate from chiral superfields and another from the anti-chiral ones.
The one-loop contributions in \cref{fig:SUSYthreshold} are given by;
\eqs{
i {\cal M}_1 & = - \frac{ig_a^2}{16\pi^2} C^{nimk} \sum_a (T^a_{jn} T^a_{ml}) F(m_{\phi_m}^2, m_{\phi_n}^2, M^2) \ave{{\cal O}_{jlik}}, \\
i {\cal M}_2 & = - \frac{ig_a^2}{16\pi^2} C^{jnlm} \sum_a (T^a_{in} T^a_{mk}) F(m_{\phi_m}^2, m_{\phi_n}^2, M^2) \ave{{\cal O}_{jlik}}, \\ 
i {\cal M}_3 & = \frac{ig_a^2}{16\pi^2} C^{nilm} \sum_a (T^a_{jn} T^a_{mk}) F(m_{\phi_m}^2, m_{\phi_n}^2, M^2) \ave{{\cal O}_{jkil}}, \\
i {\cal M}_4 & = - \frac{ig_a^2}{16\pi^2} C^{jnmk} \sum_a (T^a_{in} T^a_{ml}) F(m_{\phi_m}^2, m_{\phi_n}^2, M^2) \ave{ {\cal O}_{jkil}}. \\
}
Here, the subscript $i$ for ${\cal M}_i$ corresponds to the label $(i)$ of \cref{fig:SUSYthreshold}.
$M$ and $m_{\phi_m}$ indicate the masses of gaugino and the superpartner of $\Psi_m$, respectively.
${\cal O}_{ijkl} \equiv \ovl \Psi_i \gamma^\mu P_L \Psi_j \ovl \Psi_k \gamma_\mu P_L \Psi_l$ and $\ave{\dots}$ denotes the matrix element.
We also define the loop function $F(x,y,z)$ as 
\eqs{
F(x,y,z) = \frac34 + \frac{x^2 (y-z) \ln x + y^2(z-x) \ln y + z^2(x-y) \ln z}{2(x-y)(y-z)(z-x)} . \label{eq:loopf_susy2}
}
Here, all masses are normalized by the renormalization scale $\mu$, again.
\subsection*{In MSSM}
Let us now consider the MSSM case.
The K\"ahler potential for these operators is given by \cref{eq:dim_6}. 
The baryon-number violating four-Fermi operators are given by
\eqs{
{\cal L}_{\Del B} = \sum_{i=1}^2 C_{4F}^{(i)} {\cal O}^{(i)}_{4F}
}
with operators 
\eqs{
{\cal O}^{(1)}_{4F} & 
= \ep_{\alpha\beta\gamma} \ep_{rs} (\ovl u^\alpha \gamma^\mu P_L q^{r\gamma}) (\ovl d^\beta \gamma_\mu P_L l^s),\\
{\cal O}^{(2)}_{4F} &
= \ep_{\alpha\beta\gamma} \ep_{rs} (\ovl u^\alpha \gamma^\mu P_L q^{s\gamma}) (\ovl e \gamma_\mu P_L q^{\beta r}),
}
and Wilson coefficients $C_{4F}^{(1)} = C_{4F}^{(2)} = g_5^2/2M_X^2$.
The Wilson coefficients in the low-energy effective field theory (EFT) include threshold corrections: we need 
simply to redefine $C_{4F}^{(i)} \to (1-\lam^{(i)} )C_{4F}^{(i)} $ where $\lam^{(i)}$ denotes the threshold 
correction to $C_{4F}^{(i)}$.

After matching the amplitudes in the EFT and those in the MSSM, we find $\lam^{(i)}$ at the SUSY scale as follows;
\eqs{
\lam^{(1)} & = - \frac12 (\lam_u + \lam_q + \lam_d + \lam_l) 
- \frac{g_3^2}{16\pi^2} \left( \frac13 F(m_{\widetilde u}^2, m_{\widetilde q}^2, M_3^2) + \frac13 F(m_{\widetilde d}^2, m_{\widetilde q}^2, M_3^2) \right) \\
& - \frac{g_Y^2}{16\pi^2} \left( \frac29 F(m_{\widetilde u}^2, m_{\widetilde q}^2, M_1^2) - \frac19 F(m_{\widetilde d}^2, m_{\widetilde q}^2, M_1^2) - \frac23 F(m_{\widetilde l}^2, m_{\widetilde u}^2, M_1^2) + \frac13 F(m_{\widetilde l}^2, m_{\widetilde d}^2, M_1^2)\right), \\
\lam^{(2)} & = - \frac12 (\lam_u + \lam_e + 2 \lam_q) 
- \frac{g_3^2}{16\pi^2}  \frac23 F(m_{\widetilde u}^2, m_{\widetilde q}^2, M_3^2)  \\
& - \frac{g_Y^2}{16\pi^2} \left(\frac49 F(m_{\widetilde u}^2, m_{\widetilde q}^2, M_1^2) - \frac23 F(m_{\widetilde e}^2, m_{\widetilde q}^2, M_1^2)\right). \\
}
Here, $m_\phi ~ (\phi = \widetilde q, \widetilde u, \widetilde d, \widetilde l, \widetilde e)$ denotes the sfermions masses 
while $M_1$ and $M_3$ denote the masses of bino and gluino, respectively.
The loop function $F$ is defined in \cref{eq:loopf_susy2}.
$\lam_\psi ~ (\psi = q, l, u, d, e)$ is defined in \cref {eq:lam_psi}, and describes the one-loop threshold corrections to the two-point functions of the SM chiral fermions.

\subsection*{Numerical Evaluation}
In the last section of this appendix we evaluate the threshold effects at the SUSY scale in the split SUSY 
scenario. As mentioned in the introduction, the heavy sfermion scenario makes the constraint on a dimension-five 
proton decay mild. We assume that all sparticles except gauginos are degenerate at mass $M_S$. Gaugino 
masses are set to be as follows: the bino and wino are degenerate in mass $M_1=M_2=3~{\rm TeV}$, and we 
treat the ratio of the gluino and bino masses as a free parameter. In the numerical estimate, we set 
$M_3/M_1 = 1, 3,$ and 9. We also choose $\tan \beta = 3$ since a small $\tan \beta$ is preferred to get the 
observed Higgs mass in the heavy sfermion scenario \cite{Giudice:2004tc,Ibe:2012hu,Ibanez:2013gf}. 
The matching scale for the proton decay amplitudes is set to be the sfermion mass scale $M_S$.

\begin{figure}[t]
\centering
\includegraphics[width=7cm,clip]{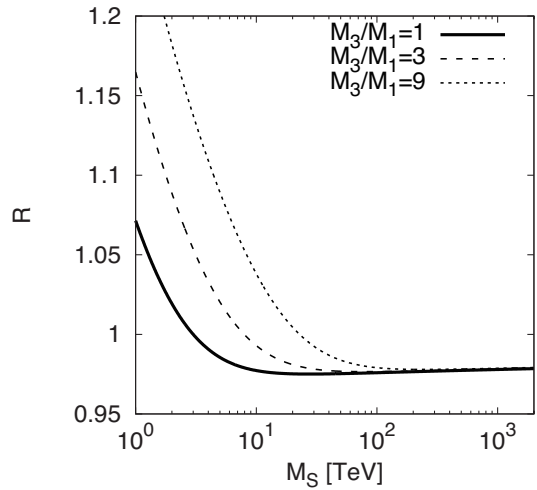}
\caption{Ratio of the proton decay rate with and  without the threshold correction at SUSY scale. At the GUT scale, we assume the minimal SUSY $SU(5)$ GUT model, and each decay rates includes the threshold correction at GUT scale.
The bino mass $M_1$ is set to be 3 TeV, and solid, broken, and dotted lines respectively indicate $M_3/M_1 = 1, ~ 3$, and 9.}
\label{fig:SUSY_threshold}
\end{figure}

We show the $M_S$ dependence of the ratio of decay rates with and without SUSY threshold correction in \cref{fig:SUSY_threshold}.
The ratio of decay rates is defined similarly as in \cref{eq:ratio}.
The denominator of \cref{eq:ratio} includes only the threshold correction at the GUT scale while the numerator also includes those at the SUSY scale. We see these threshold effects at the SUSY scale in \cref{fig:SUSY_threshold}, 
where the minimal SU(5) model with all mass parameters and the GUT scale fixed at $2 \times 10^{16} ~{\rm GeV}$ 
is considered. Let us add a few comments.

First, if all sparticles are degenerate in mass, there is no contribution from the threshold corrections.
This is because the loop functions $f$ and $F$ behave as follows ($\mu=M_S$)
\eqs{
f(M_S^2, M_S^2) \sim \frac12 \ln \frac{M_S^2}{\mu^2} \to 0, ~~~~~
F(M_S^2, M_S^2, M_S^2) \sim - \frac12 \ln \frac{M_S^2}{\mu^2} \to 0, \\ 
}

Second, the ratio $R$ approaches a constant in the limit of decoupled sfermions 
since these loop functions behave as
\eqs{
f(M_a^2, M_S^2) \sim - \frac14  \left( 1- 2 \ln \frac{M_S^2}{\mu^2}\right), ~~~~~
F(M_S^2, M_S^2, M_a^2) \sim \frac14 \left( 1 - 2 \ln \frac{M_S^2}{\mu^2}\right), \\ 
}
in the limit of $M_S \gg M_a$.  
In \cref{fig:SUSY_threshold}, the ratio $R$ is larger than 10\% when $M_3/M_1=3,~9$ and $M_S\sim O(1)$~TeV. 
This is because we take the matching scale at $M_S$ which is smaller than the gaugino masses. 
In this case we should take the matching scale somewhere between the sfermion and 
gaugino mass in order to minimize the 2-loop corrections.

As a result, we conclude that there is only a few \% correction to proton decay lifetime 
in the split SUSY scenario. 

\section{Field Decomposition \label{sec:fielddcmp}}

The K\"ahler potential for $\Phi$ in the $SU(5)$ irreducible representation is decomposed as follows; 
\eqs{
\mathcal{K} = \Phi^\dag \Phi = \sum_i \phi^\dag_i \phi_i.
}
Here, $\phi_i$ transforms as an irreducible representation under the SM gauge groups.
$\Phi^\dag$ and $\phi^\dag$ denote the anti-chiral superfields.
In this section, we turn off gauge interactions for simplicity.

In \cref{tab:decomposition}, we give the SM decomposition of the $SU(5)$ irreducible representations, 
whose Dynkin indices are below that of $\mathbf{75}$ representation. In this table, square brackets 
$[\cdots]$ represent antisymmetric indices while braces $\{ \cdots\}$ represent symmetric indices.

\begin{table}[htbp]
\caption{SM decomposition of the $SU(5)$ irreducible multiplets.
$(r_C, r_W)_{Y}$ denotes a representation transforming as $SU(3)_C$ $r_C$-plet and $SU(2)_L$ $r_W$-plet with hypercharge $Y$.}
\begin{center}
\begin{tabular}{|c|c|}
$SU(5)$ representation & labels and SM representations \\ \hline
$\Phi^A(\mathbf{5})$ &
$ \phi_1^a = (\mathbf{1},\mathbf{2})_{\frac12}, ~~~
\phi_2^\alpha = (\mathbf{3},\mathbf{1})_{-\frac13}$ \\ \hline
$\Phi^{ [AB] }(\mathbf{10})$ &
$\phi_1^{[ab]} = (\mathbf{1},\mathbf{1})_{1}, ~~~
\phi_2^{[\alpha\beta]} = (\ovl{\mathbf{3}},\mathbf{1})_{-\frac23}, ~~~
\phi_3^{a\alpha} =(\mathbf{3},\mathbf{2})_{\frac16}$ \\ \hline
$\Phi^{\{ AB\} }(\mathbf{15})$ &
$\phi_1^{ \{ ab \} } = (\mathbf{1},\mathbf{3})_{1}, ~~~
\phi_2^{\{ \alpha\beta \}} = (\mathbf{6},\mathbf{1})_{-\frac23}, ~~~
\phi_3^{a\alpha} = (\mathbf{3},\mathbf{2})_{\frac16}$ \\ \hline
\multirow{2}{*}{$\Phi^A_B(\mathbf{24})$} 
&$(\phi_1)^a_b = (\mathbf{1},\mathbf{3})_{0}, ~~~
(\phi_2)^\alpha_\beta = (\mathbf{8},\mathbf{1})_{0}, ~~~
\phi_3 = (\mathbf{1},\mathbf{1})_{0}$, \\
& $(\phi_4)^\alpha_a = (\mathbf{3},\mathbf{2})_{-\frac56}, ~~~
(\phi_5)^a_\alpha = (\ovl{\mathbf{3}},\mathbf{2})_{\frac56}$ \\ \hline
\multirow{2}{*}{$\Phi^{\{ ABC \}}(\mathbf{35})$ }
& $\phi_1^{ \{ abc \} } = (\mathbf{1},\mathbf{4})_{-\frac32}, ~~~
\phi_2^{\{ ab\} \alpha} = (\mathbf{3},\mathbf{3})_{\frac23}$ , \\
& $\phi_3^{\{ \alpha\beta \} a} = (\mathbf{6},\mathbf{2})_{-\frac16}, ~~~
\phi_4^{\{ \alpha\beta\gamma \}} = (\mathbf{10},\mathbf{1})_{-1}$ \\ \hline
\multirow{2}{*}{$\Phi^{\{ AB\} C}(\mathbf{40})$}
& $\phi_1^a = (\mathbf{1},\mathbf{2})_{\frac32}, ~~~
\phi_2^\alpha = (\mathbf{3},\mathbf{1})_{\frac23}, ~~~
\phi_3^{\{ab\} \alpha} = (\mathbf{3},\mathbf{3})_{\frac23}$, \\
& $(\phi_4)^a_\alpha = (\ovl{\mathbf{3}},\mathbf{2})_{-\frac16}, ~~~
\phi_5^{\{ \alpha\beta \}a} = (\mathbf{6},\mathbf{2})_{-\frac16}, ~~~
(\phi_6)^\alpha_\beta = (\mathbf{8},\mathbf{1})_{-1}$ \\ \hline
\multirow{3}{*}{$\Phi^{A}_{[BC]}(\mathbf{45})$} 
& $\phi_1^a  = (\mathbf{1},\mathbf{2})_{-\frac12}, ~~~
\phi_2^\alpha = (\mathbf{3},\mathbf{1})_{-\frac43}, ~~~
\phi_3^{a\alpha} = (\mathbf{3},\mathbf{2})_{\frac76}$ , \\
& $(\phi_4)^{\alpha a}_\beta = (\mathbf{8},\mathbf{2})_{-\frac12} , ~~~
(\phi_5)_{\alpha} = (\ovl{\mathbf{3}},\mathbf{1})_{\frac13}$, \\
& $(\phi_6)^a_{b\alpha} = (\ovl{\mathbf{3}},\mathbf{3})_{\frac13}, ~~~
\phi_7^{\{\alpha\beta\}} = (\mathbf{6},\mathbf{1})_{\frac13}$ \\ \hline
\multirow{2}{*}{$\Phi_{[AB][CD]}(\mathbf{50})$} 
& $\phi_1  = (\mathbf{1},\mathbf{1})_{-2}, ~~~
\phi_2^\alpha = (\mathbf{3},\mathbf{1})_{-\frac13}, ~~~
(\phi_3)^a_\alpha = (\ovl{\mathbf{3}},\mathbf{2})_{-\frac76}$ , \\
& $(\phi_4)^{[\alpha\beta]} = (\mathbf{6},\mathbf{1})_{\frac43} , ~~~
(\phi_5)_{[\alpha\beta][ab]} = (\ovl{\mathbf{6}},\mathbf{3})_{-\frac13}, ~~~
(\phi_6)_{\alpha a}^\beta = (\mathbf{8},\mathbf{2})_{\frac12}$, \\ \hline
\multirow{3}{*}{$\Phi^{\{AB\}}_C(\mathbf{70})$} 
& $\phi_1^a  = (\mathbf{1},\mathbf{2})_{\frac12}, ~~~
\phi_2^\alpha = (\mathbf{3},\mathbf{1})_{-\frac13}, ~~~
(\phi_3)^{\{ab\}}_c = (\mathbf{1},\mathbf{4})_{\frac12}$ , \\
& $(\phi_4)^{a \alpha}_b = (\mathbf{3},\mathbf{3})_{-\frac13} , ~~~
(\phi_5)^{\{ab\}}_{\alpha} = (\ovl{\mathbf{3}},\mathbf{3})_{\frac43}$, \\
& $(\phi_6)^{\alpha a}_\beta = (\mathbf{8},\mathbf{2})_{\frac12}, ~~~
(\phi_7)^{\{\alpha\beta\}}_a = (\mathbf{6},\mathbf{2})_{-\frac76}, ~~~
(\phi_8)^{\{\alpha\beta\}}_\gamma = (\mathbf{15},\mathbf{1})_{-\frac13}$ \\ \hline
\multirow{3}{*}{$\Phi^{[AB]}_{[CD]}(\mathbf{75})$} 
& $(\phi_1)_\alpha = (\ovl{\mathbf{3}},\mathbf{1})_{-\frac53}, ~~~
\phi_2^\alpha = (\mathbf{3},\mathbf{1})_{\frac53}, ~~~
\phi_3^{\alpha a} = (\mathbf{3},\mathbf{2})_{-\frac56}$ , \\
& $(\phi_4)_{\alpha a} = (\ovl{\mathbf{3}},\mathbf{2})_{\frac56}, ~~~
\phi_5^{\{\alpha\beta\}a} = (\ovl{\mathbf{6}},\mathbf{2})_{-\frac56}, ~~~
(\phi_6)_{\{\alpha\beta\}a} = (\mathbf{6},\mathbf{2})_{\frac56}$, \\
& $\phi_7 = (\mathbf{1},\mathbf{1})_{0} , ~~~ 
(\phi_8)^\alpha_\beta = (\mathbf{8},\mathbf{1})_{0}, ~~~
(\phi_9)^{\alpha a}_{\beta b} = (\mathbf{8},\mathbf{3})_{0}$ \\ 
\end{tabular}
\end{center}
\label{tab:decomposition}
\end{table}%

We summarize the vacuum polarization coefficients $b_{ij}$, defined in \cref{eq:vp_X}, from the above-mentioned 
irreducible representations. In \cref{tab:vp_X1_app}, $b_{ij}$ from each representations are shown except those 
from $\mathbf{ 5}+ \ovl{\mathbf{5}}, \mathbf{ 24} $, and $\mathbf{ 75}$ which have been already shown in \cref{tab:vp_X1}. 
As a check, the sum of $b_{ij}$ in a representation divided by its Dynkin index is a representation independent 
constant number.

\begin{table}[htbp]
\caption{Vacuum polarization coefficients $b_{ij}$.
$b_{ij}$ which are not listed here are zero.}
\begin{center}
\begin{tabular}[b]{|c|c|}
reps. & $b_{ij}$ \\ \hline
\multirow{2}{*}{$\mathbf{10}+\ovl{\mathbf{10}}$} 
& $b_{13} = b_{\ovl1 \ovl3} = 1$ \\ 
& $b_{23} = b_{\ovl2 \ovl3} = 2$ \\ \hline
\multirow{2}{*}{$\mathbf{15}+\ovl{\mathbf{15}}$} 
& $b_{13} = b_{\ovl1 \ovl3} = 3$ \\ 
& $b_{23} = b_{\ovl2 \ovl3} = 4$ \\ \hline
\multirow{3}{*}{$\mathbf{35}+\ovl{\mathbf{35}}$} 
& $b_{12} = b_{\ovl1 \ovl2} = 6$ \\ 
& $b_{23} = b_{\ovl2 \ovl3} = 12$ \\
& $b_{34} = b_{\ovl3 \ovl4} = 10$ \\ \hline
\multirow{5}{*}{$\mathbf{40}+\ovl{\mathbf{40}}$} 
& $b_{12} = b_{13} = b_{\ovl1 \ovl2} = b_{\ovl1 \ovl3} = 3/2$ \\ 
& $b_{25} = b_{35} = b_{\ovl2 \ovl5} = b_{\ovl3 \ovl5} = 3$ \\
& $b_{24} = b_{\ovl2 \ovl4} = 1/2$ \\
& $b_{34} = b_{\ovl3 \ovl4} = 9/2$ \\
& $b_{46} = b_{56} = b_{\ovl4 \ovl6} = b_{\ovl5 \ovl6} = 4$ \\ \hline
\multirow{6}{*}{$\mathbf{45}+\ovl{\mathbf{45}}$} 
& $b_{12} = b_{\ovl1 \ovl2} = 4/3$ \\ 
& $b_{15} = b_{\ovl1 \ovl5} = 1/3$ \\ 
& $b_{16} = b_{35} = b_{37} = b_{\ovl1 \ovl6} = b_{\ovl3 \ovl5} = b_{\ovl3 \ovl7} = 2$ \\ 
& $b_{24} = b_{45} = b_{\ovl2 \ovl4} = b_{\ovl4 \ovl5} = 8/3$ \\ 
& $b_{36} = b_{\ovl3 \ovl6} = 3$ \\
& $b_{46} = b_{47} = b_{\ovl4 \ovl6} = b_{\ovl4 \ovl7} = 4$ \\ \hline 
\multirow{6}{*}{$\mathbf{50}+\ovl{\mathbf{50}}$} 
& $b_{13} = b_{\ovl1 \ovl3} = 2$ \\ 
& $b_{23} = b_{\ovl2 \ovl3} = 3$ \\ 
& $b_{26} = b_{\ovl2 \ovl6} = 4$ \\ 
& $b_{35} = b_{\ovl3 \ovl5} = 6$ \\ 
& $b_{46} = b_{\ovl4 \ovl6} = 8$ \\
& $b_{56} = b_{\ovl5 \ovl6} = 12$ \\ \hline 
\multirow{8}{*}{$\mathbf{70}+\ovl{\mathbf{70}}$} 
& $b_{12} = b_{15} = b_{26} = b_{35} = b_{\ovl1 \ovl2} = b_{\ovl1 \ovl5} = b_{\ovl2 \ovl6} = b_{\ovl3 \ovl5} = 2$ \\ 
& $b_{14} = b_{\ovl1 \ovl4} = 1$ \\ 
& $b_{27} = b_{\ovl2 \ovl7} = 3$ \\ 
& $b_{34} = b_{46} = b_{\ovl3 \ovl4} = b_{\ovl4 \ovl6} = 4$ \\ 
& $b_{47} = b_{\ovl4 \ovl7} = 6$ \\ 
& $b_{56} = b_{\ovl5 \ovl6} = 8$ \\ 
& $b_{68} = b_{\ovl6 \ovl8} = 10$ \\ 
& $b_{78} = b_{\ovl7 \ovl8} = 5$ \\ 
\end{tabular}
\end{center}

\label{tab:vp_X1_app}
\end{table}%

\section{The Interaction Terms Including VEVs \label{sec:GUTvevint}}

In this appendix, we give information about the interaction terms including GUT-breaking VEVs, which are needed to calculate the vacuum polarization coefficient $a_i$ in \cref{tab:vp_X1}.
The normalization of component fields follows the definition in the previous appendix.

For an adjoint multiplet $\Phi^A_B$, the K\"ahler potential is given by
\eqs{
\mathcal{K}_{\mathbf{24}} = \Phi^{\dag A}_B (e^{-2g_5V})^C_A (e^{2g_5V})^B_D \Phi^D_C.
}
$\Phi^A_B$ obtains the GUT-breaking VEV, which is given as \cref{eq:vev24}, giving the $X$ boson a mass 
$M_X = 5  g_5 v_{\mathbf{24}}$.
The interaction terms induced by the K\"ahler potential are given by
\eqs{
\mathcal{K}_v^{\mathbf{24}} & = 2 g_5 M_X \left\{ 
- X^\dag X (\phi_1 + \phi_1^\dag) 
+ X^\dag X (\phi_2 + \phi_2^\dag) 
+ \frac{5}{\sqrt{30}} X^\dag X (\phi_3 + \phi_3^\dag) \right\} \\
& + g M_X \left\{ - G (X \cdot (\phi_4 + \phi_5^\dag))
+ W (X \cdot (\phi_4 + \phi_5^\dag))
+ \frac{5}{\sqrt{30}} B (X\cdot (\phi_4 + \phi_5^\dag)) \right\} + \text{h.c.}.
}
Here, the non-Abelian MSSM vector supermultiplets are denoted as $G \to G^\alpha_\beta = \sqrt 2 G^a (T^a)^\alpha_\beta$ and $W \to W^r_s = \sqrt 2 W^a (t^a)^r_s$, where $T^a$ and $t^a$ respectively denote the generator of $SU(3)_C$ and $SU(2)_L$.  
We also define $(A \cdot B) = \ep_{ab} A^a B^b$.

For a $\mathbf{75}$ representation $\Phi^{[AB]}_{[CD]}$, the K\"ahler potential is given by
\eqs{
\mathcal{K}_{\mathbf{75}} 
= \Phi^{\dag [AB]}_{[CD]} (e^{2g_5V})^C_E (e^{2g_5V})^D_F (e^{-2g_5V})^G_A (e^{-2g_5V})^H_B  \Phi^{[EF]}_{[GH]}.
}
The GUT-breaking VEV of $\Phi^{(AB)}_{(CD)}$ is given in \cref{eq:vev75}. The $X$ boson becomes massive 
with $M_X = 2 \sqrt 6 g_5 v_{\mathbf{75}}$ after we substitute the VEV for $\Phi$ and $\Phi^\dag$, while  
the interaction terms proportional to the VEV are
\eqs{
\mathcal{K}_v^{\mathbf{75}} & = g_5 M_X \left\{ 
 - X^\dag X (\phi_8 + \phi_8^\dag) 
+ \frac{4}{\sqrt3} X^\dag X (\phi_7 + \phi_7^\dag) 
+ \sqrt 6 X^\dag X (\phi_9 + \phi_9^\dag) \right\} \\
& - g M_X \left\{ -G (X \cdot (\phi_3 + \phi_4^\dag))
+ X (W \cdot (\phi_3 + \phi_4^\dag))
+ \frac{5}{\sqrt{30}} B ( X \cdot (\phi_3 + \phi_4^\dag)) \right\} + \text{h.c.} \\
& + \frac{\sqrt 6}{2} g M_X \ep_{ab} \ep^{\alpha\beta\gamma} X^a_\alpha X^b_\beta (\phi_1 + \phi_2^\dag)_\gamma + \text{h.c.}.
}

\newpage
\bibliography{ref}
\end{document}